\documentclass[10pt]{iopart}
\usepackage{epsf}  

\newcommand{\rthz}{\ensuremath{/\sqrt{\mathrm{Hz}}}}
\newcommand{\wsq}[1]{\ensuremath{\omega^2_{#1}}}
\newcommand{\figr}[1]{Fig. \ref{#1}}
\newcommand{\eqr}[1]{Eqn. \ref{#1}}

\newcommand{\Sh}[1]{$S_{#1}^{1/2}$}

\begin{document}

\title[]
{Measuring random force noise for LISA aboard the 
LISA Pathfinder mission}

\author{
D.~Bortoluzzi,$^a$ L.~Carbone,$^b$ A.~Cavalleri,$^c$ 
M.~Da~Lio,$^a$ R.~Dolesi,$^b$ C.~D.~Hoyle,$^b$ 
M.~Hueller,$^b$ S.~Vitale,$^b$ and W.~J.~Weber$^b$
}

\address{$^a$ Dipartimento di Ingegneria, 
Universit\`{a} di Trento, 38100 Trento, Italy}
\address{$^b$ Dipartimento di Fisica and INFN,
Universit\`{a} di Trento, 38050 Povo, Trento, Italy}
\address{$^c$ Centro Fisica degli Stati Aggregati, 38050 Povo, Trento, Italy}

\ead{weber@science.unitn.it}

\begin{abstract}
The LTP (LISA Testflight Package), to be flown aboard the
ESA~/~NASA~LISA Pathfinder mission, aims to demonstrate drag-free
control for LISA test masses with acceleration noise below 30~fm
/s$^2$\rthz\ from 1-30~mHz. This paper describes the LTP measurement of
random, position independent forces acting on the test masses. In
addition to putting an overall upper limit for all source of random
force noise, LTP will measure the conversion of several key disturbances
into acceleration noise and thus allow a more detailed characterization
of the drag-free performance to be expected for LISA. 
\end{abstract}

\pacs{4.80.Nn}



\section{Introduction}
The gravitational wave sensitivity for the Laser Interferometer Space
Antenna (LISA) will be limited at low frequencies 
by the stray acceleration noise in the
orbits of the nominally free-falling test masses that serve as
interferometry end mirrors. The test mass (mass $m$) acceleration noise
$a_n$ is typically divided into a contribution from random, position
independent forces ($f_{str}$) and another from coupling (with spring
constant $m \wsq{p}$) to the noisy motion of the spacecraft shield:
\begin{equation}
a_{n} = \frac{f_{str}}{m} + \wsq{p} 
\left( x_n + \frac{F_{str}}{M \wsq{DF}} \right)  
\label{eqn_LISA}
\end{equation}
The closed loop satellite position noise arises in the noise $x_n$ of
the position sensor used to guide the satellite control and in the
imperfect compensation of external forces $F_{str}$ acting on the
satellite by the finite gain thruster control loop (gain $\wsq{DF}$).
LISA aims to limit the test mass acceleration noise spectral density to
$S_{a_n}^{1/2} < 3 \times 10^{-15} \, \mathrm{m / s}^2\rthz$ (3
fm/s$^2$\rthz) at frequencies down to 0.1~mHz \cite{big_book}. 

LTP is a single spacecraft experiment that tests drag-free control for
LISA by measuring the differential motion of two test masses, each
``free-falling'' inside a LISA capacitive position sensor, along a
single measurement axis (the LTP configuration and measurement schemes
are described in Ref. \cite{cqg_ltp}, and a schematic of the apparatus
design is shown in \figr{LTP_figure}). The simpler 1-spacecraft,
1-axis configuration requires use of control forces
along the measurement axis, a performance limiting departure from LISA.
The main measurement for LTP, in which the satellite is controlled to
follow the first test mass (TM1) while the second test mass (TM2) is
forced electrostatically to follow the satellite, aims to put an overall
acceleration noise upper limit of 30~fm/s$^2$\rthz\ at 1~mHz, relaxed by
an order of magnitude with respect to LISA in both noise level and
frequency. 

In addition to a global limit on acceleration noise, LTP will fully
characterize the satellite coupling term in \eqr{eqn_LISA}. The external
force level $F_{str}$ can be extracted from the closed loop position
sensor error signal, and the sensor noise can be obtained by comparison
with a more precise optical readout. ``Stiffness'' \wsq{p} is measured
by modulating the satellite control setpoint \cite{cqg_ltp}. 

\begin{figure}
\begin{center}
\epsfbox[13 14 390 227]{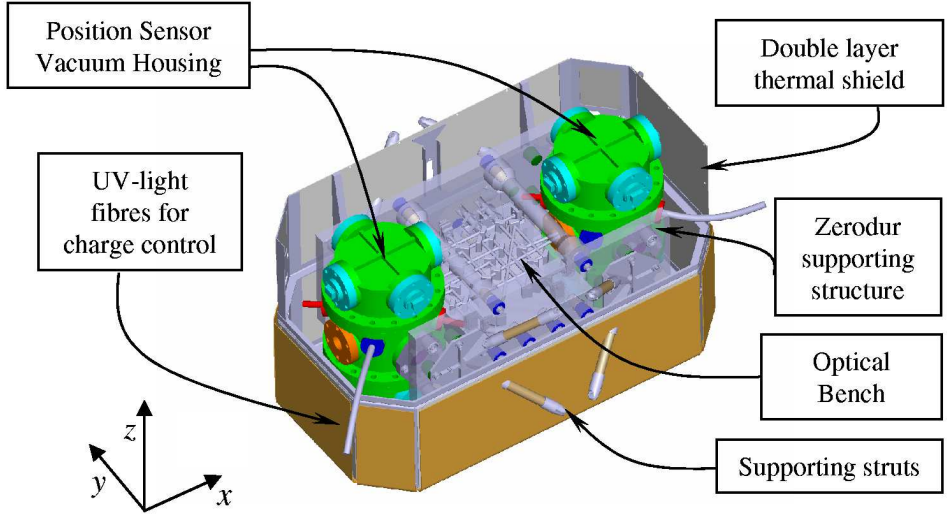}
\epsfxsize=5in
\end{center}
\caption{\label{LTP_figure} Design of the LTP experimental apparatus,
currently at an engineering model phase of development and construction.
The heart of the apparatus is the two test masses, each one flying inside
a capacitive position sensor, with an optical interferometer measuring their
differential displacement along the $x$ axis.  Other papers within this
proceedings address the optical interferometry \cite{g_money_amaldi_5} 
and UV light charge management system \cite{t_money_amaldi_5}.
}
\label{LTP_figure}
\end{figure}

This paper focuses on the random force measurements $f_{str}$ that will
be possible aboard LTP. While the overall acceleration noise limit
possible with LTP is an order of magnitude above the LISA goal, a
special control scheme isolates the random force contribution, allowing
measurement of \Sh{f_{str}}\ to within a factor 2 of the LISA goal for 
most sources.
Additionally, several critical noise sources can be modulated coherently
for precise measurement of their coupling into test mass acceleration.
Combined with the measurements of \wsq{p}, $x_n$, and $F_{str}$ discussed
above, the random force measurement pushes the characterization of the
LISA noise budget to lower acceleration noise levels.
Precise measurement of key noise parameters will allow dissection of the
random noise measurement as well as a more accurate extrapolation of the
noise model to the lower frequencies and lower disturbance levels needed
for LISA. 

\section{LTP random force noise measurement}
\begin{figure}
\begin{center}
\epsfbox[89 82 465 341]{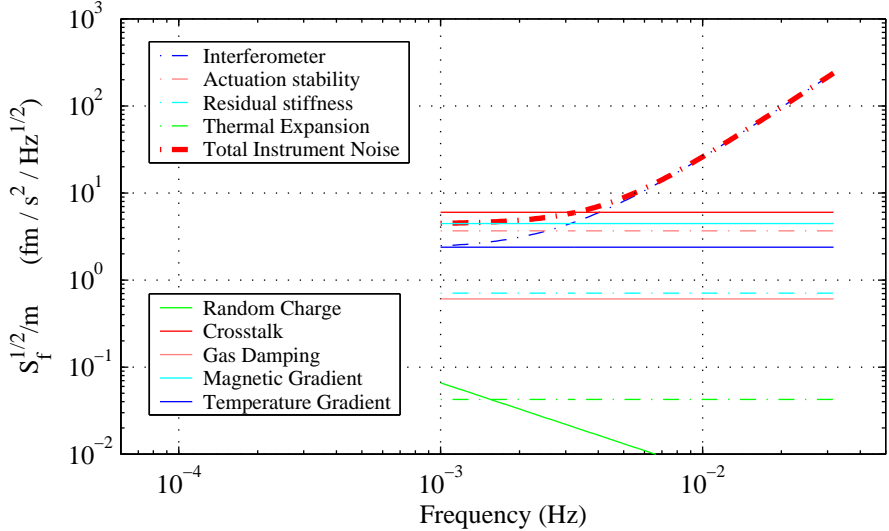}
\epsfxsize=3in
\end{center}
\caption{\label{LTP_instr_noise} Plot of key instrumental noise sources
(dashed) for the LTP random force measurement, shown with key random
force sources (solid) as they could appear for LTP. Curves are
normalized to noise on a single test mass and assume uncorrelated forces
acting on the two test masses.}
\label{LTP_instr_noise}
\end{figure}
The control scheme for the LTP random force measurement is a modified
version of that used for the overall acceleration noise measurement
mentioned above. The satellite is controlled to follow TM1, and TM2 is
controlled, by nulling the differential displacement interferometric
readout, to follow TM1. In this configuration, the residual error signal
measured by the interferometer can be expressed 
\begin{eqnarray} 
\label{eq_LTP_m3}
\fl 
\Delta x_{opt} = 
\frac{1}{\wsq{} - \left( \wsq{2p} + \wsq{ES} \right) } \times 
\\ \nonumber 
\fl \: \: \: \: \: \: \: \: \: \: \: \: \: \: \:
\left[
 \frac{f_1 - f_2}{m}
+ \left( x_{n1} + \frac{F_{str}}{M \omega^2_{DF}} \right) 
 \left[ \wsq{1p}  -  \wsq{2p} \right] 	 
- \delta x \: \wsq{2p}
+ x_{n,opt} \left( \wsq{} - \wsq{2p} \right)
\right]
\end{eqnarray}
Indices 1 and 2 refer to TM1 and TM2, $x_{n,opt}$ is the differential
interferometry noise, $\delta x$ is the distortion of the baseline
separating the two position sensors, and \wsq{ES}\ is the electrostatic
control loop gain for TM2. Here the satellite motion couples to the
differential interferometry signal only through the stiffness difference
$\left[ \wsq{1p} - \wsq{2p} \right]$, a small quantity which can be
measured and, if necessary, tuned electrostatically to negligible
magnitude \cite{cqg_ltp}. This control scheme thus minimizes the effect
of stiffness, which is the limiting factor in the overall acceleration
noise measurement, and thus isolates the random forces $f_1$ and $f_2$.

The noise in the interferometry signal $\Delta x_{opt}$ in
\eqr{eq_LTP_m3} is thus a measurement of $f_1 - f_2$ for LISA, with
several additional LTP specific ``instrumental'' noise sources. In
addition to the negligible contribution from satellite coupling, the
baseline distortion term ($\delta x$) is also negligible for the high
stability Zerodur optical bench and 10$^{-4}$~K\rthz\ temperature
stability projected for LTP. The remaining dominant instrumental noise
sources for the LTP measurement of random force noise are then: 

\begin{itemize}
%

\item{\it{Interferometry noise }} 
Noise in the differential interferometry readout converts into an effective
random force noise (directly from \eqr{eq_LTP_m3})
\begin{equation}
\frac{S^{1/2}_{f_{opt}}}{m} = 
\frac{1}{\sqrt{2}} \, S^{1/2}_{x_{n,opt}} \left| \wsq{} - \wsq{2p} \right|
\label{eqn_interfere}
\end{equation}
The LTP interferometry requirement is roughly 8~pm\rthz\ above~3 mHz and
is relaxed as $\frac{1}{f^2}$ at lower frequencies \cite{gerhard}. This
interferometry noise also includes a contribution from the thermal
expansion of the test masses and optical windows, though this term
should be small. Interferometry noise is the dominant instrumental noise
source above 3 mHz. 

\item{\it{Actuation noise }} 
Any instability in the applied electrostatic actuation forces 
produces acceleration noise given by \cite{SPIE_sens}
\begin{equation}
\frac{S^{1/2}_{f_{ACT}}}{m} = 
\frac{1}{\sqrt{2}} \, \frac{S^{1/2}_{F_{ACT}}}{m} = 
\frac{2}{\sqrt{2}} \, \Delta a_{DC} \, S^{1/2}_{\Delta V / V}
\label{eqn_actuation}
\end{equation}
While actuation noise is part of the random force $f_2$, we include it
as an LTP instrumental noise source because it originates in the need to
compensate a DC acceleration imbalance ($\Delta a_{DC}$) in LTP's 1-axis
measurement configuration and is not present in LISA. Limiting this
noise source requires both high actuation voltage stability levels
(relative fluctuations \Sh{\Delta V/V} $\approx 2 \times 10^{-6}$ \rthz\
at 1~mHz) and tight gravitational balancing of the satellite ($\Delta
a_{DC}~\approx$~1.3~nm/s$^2$). This is likely the dominant instrumental
limitation to the LTP random force noise measurement below 3~mHz. 
\end{itemize}

Here we have used \eqr{eq_LTP_m3} to convert these instrumental
disturbances into effective force noise, normalized to the random force
noise on a single test mass \Sh{f_{str}} and assuming that $f_1$ and $f_2$
are uncorrelated. The instrument noise limit for measuring such
uncorrelated noise sources is roughly 5~fm/s$^2$\rthz at 1~mHz, 
within a factor 2
of the LISA goal. While the uncorrelation assumption holds for many
sources, including electrostatic effects, circuitry back-action,
and Brownian noise sources, it is less valid for noise from temperature
and magnetic field noise, which can act coherently on the two test
masses (the instrumental noise source conversion factor $\sqrt{2}$ in
the denominator of Eqns. \ref{eqn_interfere} and \ref{eqn_actuation}
would range from 0-2 depending on the nature and degree of correlation).
We should point out, however, that these environmental disturbances are
dominated by short range, on-board sources, and, additionally, that the 
physical mechanisms coupling to the environment (such as the random residual 
magnetic moment) are not likely to be equal in the two test masses.  
As such, a substantial cancellation, which would render LTP insensitive to
such effects, is unlikely..

Before addressing individual random force noise sources relevant to LTP
and LISA, we should clarify the frequency dependence of the sources in
\figr{LTP_instr_noise}. First, noise sources which are not inherently
frequency dependent (for thermal gradient effects, for example,
$S^{1/2}_{f_{th}} \sim S^{1/2}_{\Delta T}$) are drawn as flat
(``white'') noise sources, even though they are likely to have a
``pink'' spectral shape due to a very likely low frequency increase in
the driving noise source (temperature fluctuations in this example). In
the absence of detailed spectral information, we choose to apply here
the low frequency (1~mHz for LTP, 0.1~mHz for LISA) target for the given
fluctuations (thermal, magnetic, sensor noise, etc) even at higher
frequencies where one would expect the noise to improve. For the same
reason, we extend the LTP noise curves only to 1 mHz, the official low
frequency limit of the mission. LTP will measure to even lower
frequencies, but with sensitivity likely deteriorating as
$\frac{1}{f^2}$ or worse, considering both the actuation and
interferometer performance. 

\section{Characterizing specific LISA stray force sources with LTP}
Several key random force noise sources are plotted in
\figr{LTP_instr_noise} for LTP and in \figr{LISA_noise} for LISA
(References
\cite{SPIE_sens,rita_LISA_symp,arch,hanson_LISA_symp,bonny_LISA_symp}
all address noise sources for LTP, DRS, and LISA). The LTP random force
noise measurement can, for specific known disturbances, be better
extrapolated to the frequencies and environmental conditions of LISA by
making precise measurements of the parameters that govern these noise
sources. Additionally, coupled with simultaneous measurements of
environmental effects, they also permit debugging of the LTP force
measurement itself, through correlation and possible subtraction of
disturbances from the interferometry time series. This will also allow
characterization of a noise source that might have escaped the random
noise measurement because of a coherent cancellation between the two
test masses. 

To characterize a specific noise source, we employ the differential
interferometer to detect forces excited by coherently modulating the
given disturbance (on one or both test masses), with the low background
force noise allowing fN measurement resolution in a 1~hour measurement
\cite{cqg_ltp}. Envisioned measurements of known sources include:

\begin{figure}
\begin{center}
\epsfbox[89 82 465 341]{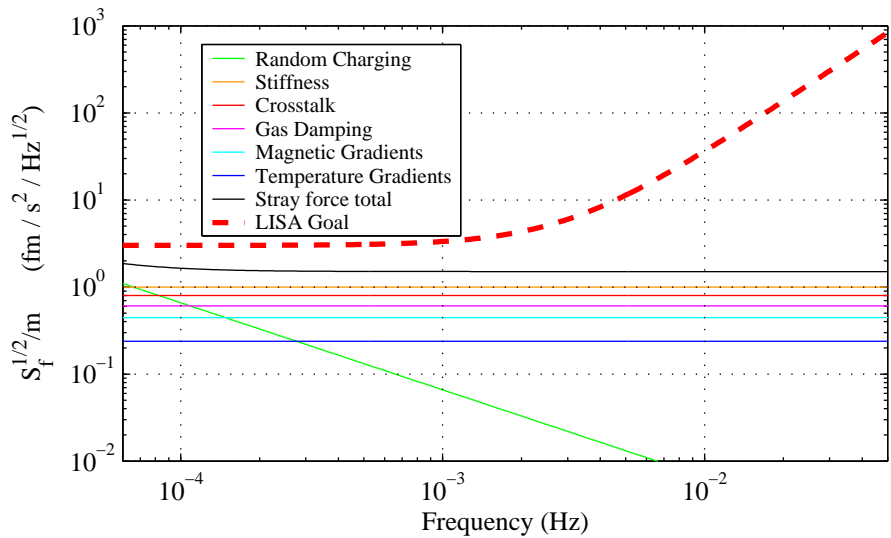}
\epsfxsize=5in
\end{center}
\caption{\label{LISA_noise} Plot of key acceleration noise sources 
for LISA, including
the satellite coupling (``stiffness'') that is insignificant in the
LTP random force measurement.  }
\label{LISA_noise}
\end{figure}


\begin{itemize}
\item{\it{Magnetic moment }}
Interaction of the test mass magnetic moment ($\vec{m}$) with magnetic
field ($\vec{B}$) fluctuations produce a force $F_x
=\frac{\partial}{\partial x} \left( \vec{m} \cdot \vec{B} \right)$, with
moment $\vec{m}$ likely dominated by the remnant ferromagnetic moment
$\vec{m_0}$, which should be below 0.02~$\mu$A m$^2$ for the Au - Pt
test mass. The field gradient goal for LTP is 0.25~$\mu$T/m\rthz,
improving to 0.025~$\mu$T/m\rthz\ for LISA. LTP will have magnetometers
and Helmholtz coils to measure and apply magnetic fields. With two coils
oriented along the $x$-axis, symmetrically on either side of a test
mass~/~sensor, the moment $m_{0x}$ can be measured by modulating a
magnetic gradient $\frac{\partial B_x}{\partial x}$ (currents opposing
in the two coils) and measuring the resulting force in $x$. Components
$m_y$ and $m_z$ can be measured by applying a homogeneous field along
$x$ (parallel coil currents) and observing the torques excited about,
respectively, the $z$ and $y$ axes. In both cases field levels of order
10~$\mu$T are sufficient for moment measurement at the percent level.
While the test mass magnetic moment will also be measured on the ground,
measurement in-flight will check for possible magnetic contamination
occurring during final preparations, launch, or flight. 

\item {\it{Temperature gradient effects}}
Temperature differences between opposing faces in the position sensors
that surround the test masses create forces through the radiometric
effect and differential radiation pressure. For the 10$^{-5}$~Pa
pressure projected for the LTP position sensor vacuum chambers,
differential radiation pressure is likely to be roughly twice as large
as the radiometric effect, producing 2 fm/s$^2$\rthz\ for the envisioned
1~mHz temperature difference stability of 10$^{-4}$~K\rthz\ (LISA aims
to improve this thermal stability by a factor 10 down to 0.1~mHz). An
additional, less predictable thermal gradient force disturbance could
arise in any temperature dependent outgassing of the sensor walls. To
measure the total temperature gradient ``feedthrough,'' LTP will be
equipped with thermometers and heaters, to excite and measure
temperature differences across the position sensors, with differences as
small as 10~mK allowing measurement within 1\% of the radiation pressure
effect. 

\item{\it{Stray DC electrostatic fields}}
Stray DC biases on the electrode surfaces of the capacitive position
sensors can interact with the noisy test mass charge and dielectric
noise to produce low frequency acceleration noise \cite{cospar_DC}. The
DC biases can be measured with modulated sensing voltages and then
compensated by applying appropriate ``counter-bias'' voltages with the
actuation circuitry. Measurement of stray DC biases of tens of mV, and
their compensation to within 1~mV, has been demonstrated on ground
\cite{pend_prl} and will be performed on LTP and LISA to significantly
reduce this potentially dangerous noise source (Figs.
\ref{LTP_instr_noise} and \ref{LISA_noise} assume a more modest
compensation to within 10~mV). Longer measurements could also
characterize the noise in the stray bias itself, a possible noise
source at very low frequencies. 

\item{\it{Cross-talk effects}}
The coupling of residual satellite motion along the non-measurement
translational and rotational degrees of freedom into acceleration along
the critical $x$ axis is an important and complicated noise source for
both LTP and LISA. One example is the gravity gradient $\frac{\partial
g_x}{\partial z}$, which is dominated by the massive optical bench lying
just beneath the axis connecting TM1 and TM2 and could be as large as $2
\times 10^{-8}$/s$^{-2}$; this accelerates a test mass in $x$ in
response to satellite motion in $z$. Another example is the $x$
acceleration noise arising through the slight rotation of the $z$ and
$y$ actuation forces with the satellite rotational noise. Cross-talk
effects could be as big as 5-10~fm/s$^2$\rthz\ in the LTP random force
noise measurement, where a modest level of off-axis satellite control is
considered acceptable for cost reasons (LTP requires 70~nm\rthz\ on the
$y$ and $z$ axes, compared to 10 for LISA). Cross-talk feedthrough
coefficients can be measured by coherent modulation, at the $\mu$m
level, of the spacecraft position via the control setpoints on different
axes.

\end{itemize} 

\section{Conclusions}
Achievement of LTP's principle scientific objective, demonstrating an
overall acceleration noise limit of 30~fm/s$^2$\rthz\ at 1~mHz, is
sufficient to establish LISA's capability to detect a host of
interesting gravitational wave sources \cite{big_book}. The higher
precision measurement of the random force noise discussed here, to
within roughly twice the LISA goal at 1~mHz, will further increase
confidence in LISA's performance goals. This improves substantially upon
the resolution of earth based torsion pendulum measurements of stray
forces (currently roughly two orders of magnitude above LISA's goal at 1
mHz \cite{pend_prl}) and confronts many of the challenges of precision
metrology in space, from launch survival to all-axes satellite control,
that LISA will face. Additionally, the coherent disturbance experiments
will extend modelling of key noise sources to LISA's lower frequencies,
with noise parameter measurements performed in representative flight
conditions subject to possible damage and contamination, from handling,
launch, or the satellite environment, that could affect sensitive
surface electrostatic, magnetic, or outgassing properties for LISA.

\end{document}